\documentclass[11pt]{article}
\pdfoutput=1
\usepackage[top=2cm, bottom=2cm, left=2cm, right=2cm]{geometry}
\usepackage{jcappub}
\usepackage{float}
\usepackage{graphicx}
\usepackage{rotating}
\usepackage{color}
\usepackage{amsmath,bm}

\newcommand{\sv}{\langle\sigma v \rangle}

\newcommand{\D}{\mathrm{d}}

\newcommand{\diff}{\mathrm{d}}

\definecolor{purple}{RGB}{160,0,160}
\definecolor{plotpink}{RGB}{205,0,180}
\definecolor{plotcyan}{RGB}{0,215,215}
\definecolor{plotblue}{RGB}{0,0,235}
\definecolor{plotorange}{RGB}{245,140,0}
\definecolor{plotgreen}{RGB}{30,130,0}
\definecolor{plotred}{RGB}{240,0,0}
\definecolor{darkgreen}{RGB}{0,170,0}

\title{Constraining heavy dark matter with cosmic-ray antiprotons}
\author[a]{Alessandro Cuoco,}
\author[a]{Jan Heisig,}
\author[a,b,c]{Michael Korsmeier}
\author[a]{and Michael Kr\"amer}
\affiliation[a]{Institute for Theoretical Particle Physics and Cosmology,
RWTH Aachen University, 52056 Aachen, Germany}
\affiliation[b]{Dipartimento di Fisica, Universit\`a di Torino, via P. Giuria 1, 10125 Torino, Italy}
\affiliation[c]{Istituto Nazionale di Fisica Nucleare, Sezione di Torino, Via P. Giuria 1, 10125 Torino, Italy}

\emailAdd{cuoco@physik.rwth-aachen.de}
\emailAdd{heisig@physik.rwth-aachen.de}
\emailAdd{korsmeier@physik.rwth-aachen.de}
\emailAdd{mkraemer@physik.rwth-aachen.de}

\abstract{
Cosmic-ray observations provide a powerful probe of dark matter annihilation
in the Galaxy.
In this paper we derive constraints on heavy dark matter from the recent precise AMS-02 antiproton data.
We consider all possible annihilation channels into pairs of
standard model particles. Furthermore, we interpret our results in the context of minimal dark matter, 
including higgsino, wino and quintuplet dark matter. 
We compare the cosmic-ray antiproton limits to limits from $\gamma$-ray observations of dwarf spheroidal galaxies and to limits from $\gamma$-ray and $\gamma$-line observations towards the Galactic center.
While the latter limits are highly dependent on the dark matter density distribution and only exclude a thermal
wino for cuspy profiles, the cosmic-ray limits are more robust, strongly disfavoring the thermal wino dark matter scenario even for 
a conservative estimate of systematic uncertainties.}

\subheader{\hfill \textnormal{TTK-17-34}}

\arxivnumber{}

\begin{document}

\maketitle
\flushbottom

%===================================================================
\section{Introduction}\label{sec:intro}
%===================================================================

Indirect detection with antiprotons is an important search strategy to test the
dark matter (DM)  self-annihilating nature 
predicted by thermal freeze-out scenarios \cite{Bergstrom:1999jc,Donato:2003xg,Bringmann:2006im,Donato:2008jk,Fornengo:2013xda,Pettorino:2014sua,Bringmann:2014lpa,Cirelli:2014lwa,Hooper:2014ysa,Cembranos:2014wza,Boudaud:2014qra,Giesen:2015ufa,Jin:2015sqa,Evoli:2015vaa}.
Thanks to recently published precise AMS-02 cosmic-ray (CR) antiproton 
data \cite{Aguilar:2016kjl}  a significant reduction of the uncertainties
related to the CR propagation is possible, thus providing stringent 
DM constraints \cite{Cuoco:2016eej,Cuoco:2017rxb}.
Compatible results has been also found when following the usual strategy of using Boron over Carbon data (B/C) 
to constrain the propagation scenario \cite{Cui:2016ppb}.
Nonetheless, determining the CR propagation directly from antiproton
data as opposed to B/C data 
has the advantage of avoiding possible biases from the possibility that heavier nuclei like B,C propagate differently from lighter ones such as p, He, $\bar{p}$.
A disadvantage is, however, that DM and propagation have to be fitted
together since they both affect the antiproton spectrum, and
correlations and degeneracies among the two are possible.
This complication can nonetheless be taken into account  \cite{Cuoco:2016eej,Cuoco:2017rxb}.

The analyses in  \cite{Cuoco:2016eej,Cuoco:2017rxb,Cui:2016ppb} have mainly focused
on the DM mass range around 100 GeV, since this range provides an interesting hint
of a signal compatible with the presence of a DM particle annihilating
with a thermal cross section. At larger masses, instead, stringent upper limits on the 
annihilation cross-section have been set, although only for a 
limited choice of annihilation channels, specifically $b\bar{b}$ and $W^+W^-$.
In this paper we extend the
study of~\cite{Cuoco:2016eej,Cuoco:2017rxb} in two ways. On the one hand,
we derive upper limits including all possible annihilation channels into pairs of
Standard Model (SM) particles. We also improve the methodology employed to derive the limits. 
Despite the fact that the sensitivity for leptonic 
channels is not as strong as for annihilation into quarks, gluons, gauge or Higgs bosons, a substantial production
of antiprotons is possible also for leptonic channels at large DM masses
thanks to electroweak corrections. We will show that for some of these channels,
and for large  DM masses, indeed the antiproton constraints are competitive
or stronger than the ones derived from gamma-ray observations.
Secondly, we will derive upper limits for specific DM models, namely minimal DM~\cite{Cirelli:2005uq} which extends the SM by the inclusion of electroweak multiplets, 
and which requires DM masses in the TeV-range to provide the observed DM relic abundance.
In particular, we find that the thermal triplet model (wino DM) is strongly disfavoured, even when taking 
into account a conservative estimate of the systematic uncertainties in the antiproton limits.

The paper is organized as follows. In section~\ref{sec:limits} we present our general analysis of DM bounds from comic ray antiprotons, as well as from gamma-ray observations of dwarf spheroidal galaxies and gamma-ray observations towards the Galactic center. Specific minimal DM models, including wino, higgsino and fermion quintuplet DM are introduced in section~\ref{sec:models}. We probe these models with cosmic ray antiprotons and compare our limits with results obtained from $\gamma$-line searches. Numerical results for the various models are presented in section~\ref{sec:MDMresults}. We conclude in section~\ref{sec:summary}. 

%===================================================================
\section{Dark matter limits}\label{sec:limits}
%===================================================================

In this section we derive generic DM limits in a model-independent way 
considering all possible annihilation channels into pairs of SM particles.
Limits for specific models will be discussed in section~\ref{sec:models}.

%------------------------------------------------------------------------
\subsection{Cosmic-ray analysis}
%------------------------------------------------------------------------

Antiprotons can be produced through the fragmentation of the products of DM annihilation in the Galaxy.
This corresponds to a source term given by
\begin{eqnarray}
  \label{eqn::DM_source_term}
  q_{\bar{p}}^{(\mathrm{DM})}(\bm{x}, E_\mathrm{kin}) = 
  \frac{1}{2} \left( \frac{\rho(\bm{x})}{m_\mathrm{DM}}\right)^2  \sum_f \left\langle \sigma v \right\rangle_f \frac{\diff N^f_{\bar{p}}}{\diff E_\mathrm{kin}} ,
\end{eqnarray}
where  $\rho(\bm{x})$ is the DM density distribution, $m_\mathrm{DM}$ is  the  DM mass,
$\left\langle \sigma v \right\rangle_f$ is the  thermally averaged cross section for DM annihilation into the SM final state $f$, ${\rm DM+DM} \to f\bar{f}$, and
$\diff N^f_{\bar{p}}/\diff E_\mathrm{kin}$ is the antiproton 
energy spectrum per DM annihilation. The factor $1/2$ in eq.~\eqref{eqn::DM_source_term} has to be included for scalar or Majorana fermion DM\@.
For the analysis, we consider different DM distributions.
As default, we use the NFW DM density profile~\cite{Navarro:1995iw}, 
 \begin{equation}
   \rho_{\mathrm{NFW}}(r) =   \frac{\rho_\text{h}}{(r/r_\text{h})(1 + r/r_\text{h})^2}\,,
 \end{equation}
with a halo scale radius of $r_\text{h}=20\,$kpc, and   a halo density, $\rho_\text{h}$, 
normalized to the local DM density $\rho_\odot = 0.43\,$GeV/cm$^3$~\cite{Salucci:2010qr} 
at the solar position, $r_ \odot = 8\,$kpc. 
Furthermore, we consider the Burkert profile~\cite{Burkert:1995yz}, 
 \begin{equation}
  \rho_{\mathrm{Bur}}(r) =  \frac{\rho_\text{c}}{(1+r/r_\text{c}) (1 + r^2/r_\text{c}^2 )}\,,
 \end{equation}
with a core radius of $r_c=5\,$kpc and $r_c=10\,$kpc, again normalized to give $\rho_\odot = 0.43\,$GeV/cm$^3$ at the solar position.
In both the NFW and Burkert case the parameters have been chosen in agreement with
the recent determination of the DM halo in \cite{Nesti:2013uwa}.

The yield and energy distribution of antiprotons per DM annihilation, ${\diff N^f_{\bar{p}}}/{\diff E_\mathrm{kin}}$, depends on the SM final state $f$ and the DM mass. We consider the channels 
$f=q, c, b, t, g, W,$ $Z, h, \ell$ and $\nu$, where $q=u,d,s$, $\ell=e,\mu,\tau$ and  $\nu=\nu_e,\nu_\mu,\nu_\tau$, and use 
the results for ${\diff N^f_{\bar{p}}}/{\diff E_\mathrm{kin}}$ presented in \cite{Cirelli:2010xx}. 
The authors of  Ref.~\cite{Cirelli:2010xx} took into account electroweak corrections~\cite{Ciafaloni:2010ti}
in a model-independent way by using electroweak splitting functions~\cite{Ciafaloni:2001mu,Ciafaloni:2005fm}.
This approximation is justified for final states with energies well above the mass of the weak gauge bosons and has been shown to reproduce full matrix element computations for $m_\text{DM}\gtrsim500\,\text{GeV}$~\cite{Cavasonza:2014xra}. Note that for the leptonic channels antiprotons arise solely from electroweak radiation. 

We analyze the effect of DM annihilation on the CR antiproton flux by performing a combined fit 
of the fluxes of protons, helium and antiprotons including the contribution from DM annihilation in the Galaxy,
 eq.~\eqref{eqn::DM_source_term}, as an additional source term. 
We use  AMS-02 proton and helium fluxes \cite{Aguilar_AMS_Proton_2015, Aguilar_AMS_Helium_2015}, 
and the AMS-02 antiproton to proton ratio \cite{Aguilar:2016kjl}, together with 
proton and helium data from CREAM~\cite{Yoon_CREAM_CR_ProtonHelium_2011} and VOYAGER~\cite{Stone_VOYAGER_CR_LIS_FLUX_2013}
We use \textsc{Galprop} \cite{Strong:1998fr,Strong:2015zva} to solve the standard CR propagation equation.
\textsc{Galprop} is run in cylindrical symmetry mode, and with the Galaxy radial extension fixed to 20 kpc.
 The propagation is parametrized by a total of 16 parameters,
 which we scan using \textsc{MultiNest} \cite{Feroz_MultiNest_2008}. Six parameters are used to describe the injection spectrum of protons and helium. In addition, there are the normalization and slope of the diffusion coefficient,  $D_0$ and $\delta$, respectively, 
 the velocity of Alfven magnetic waves, $v_A$, the convection velocity, $v_{0c}$, 
 and the Galaxy's half-height, $z_h$.
 Two parameters, $m_{\rm DM}$ and  $\sv$, characterize the contribution of DM annihilation.
 Three more parameters, the normalization of the proton and helium fluxes, $A_\mathrm{p}$ and $A_\mathrm{He}$, respectively, 
 and the solar modulation potential, $\phi_\mathrm{AMS}$, can be varied
 without reevaluating \textsc{Galprop},  reducing the parameter space to be explored to effectively 13 dimensions. For more details on the methodology we  refer to~\cite{Korsmeier:2016kha,Cuoco:2016eej}. 
 
In our setup the background for DM searches is thus given by secondary antiprotons
 produced by primary proton and helium CRs during propagation.   
We do not consider possible contributions to the background by 
primary antiprotons which could be produced directly in the astrophysical CR sources
and that would contribute mainly above $\sim$ 100 GV  (see for example \cite{Blasi:2009hv,Blasi:2009bd,Fujita:2009wk,Kohri:2015mga,Cholis:2017qlb}).
In fact, in our work we find that secondary antiprotons
fit well the measured spectrum also at high energies above 100 GV.
Such a primary contribution is thus not required in our analysis.
Furthermore, such a component would be basically degenerate with 
primary antiprotons from  heavy DM (mass $\gtrsim$ 1 TeV).
Not including it, would thus produce, eventually, more conservative
DM limits, since DM would `absorb' the primary astrophysical component.

As benchmark production cross section for the astrophysical background of  antiprotons
we use the model presented in~\cite{Mauro_Antiproton_Cross_Section_2014} for the $pp$-induced production, and the scaling discussed in~\cite{Donato:2017ywo} for the production which involves He.
We quantify the dependence of our DM limits on
the antiproton cross section by using alternative parametrizations such as~\cite{TanNg_AntiprotonParametrization_1983} (default in \textsc{Galprop}), as well as \cite{Kachelriess:2015wpa,Kappl:2014hha,Winkler:2017xor}.
The very recent measurement of antiproton production in proton-helium scattering by LHCb~\cite{Graziani:2017xas} is described well by the cross-section parametrizations \cite{Donato:2017ywo} and, in particular, \cite{Winkler:2017xor}. Note, however, 
that the LHCb measurement has only little impact on the prediction of the antiproton flux for AMS-02 energies~\cite{Reinert:2017aga, Korsmeier:2018gcy}.

Previous analyses of the antiproton spectrum~\cite{Cuoco:2016eej,Cui:2016ppb,Cuoco:2017rxb}   have revealed a potential DM signal with DM masses in the range between about 50 and 200~GeV, depending in detail on the annihilation channel. In the present analysis we focus on heavy DM, with masses between 200~GeV and 50~TeV, and derive limits on the $2\to 2$ DM annihilation cross section, $\sv_f$, for different SM final states $f$. 
We use a  frequentist  procedure and construct the profile likelihood as a function
of $\sv$ for a given DM mass, marginalized over the parameters of the cosmic ray propagation.
To estimate the systematic uncertainties associated with the DM limits, we repeat the analysis 
using different setups. In particular, we vary the diffusion model considering a case without convection,
and cases with a fixed $z_h$, we use different DM profiles, different rigidity ranges in the fit, and different models of antiproton production cross-sections, see the discussion in section~\ref{sec:resdis}

To avoid performing different scans for each DM mass, however, we use the results
of the global scan where $m_{\rm DM}$ and  $\sv$ are varied.
In practice, we divide the mass range from 200 GeV to 50 TeV into 20 logarithmically spaced
bins, and construct the profile likelihood in each bin from the likelihood samples collected in the scan.
Limits are then set at 95\% C.L. from the condition $\Delta \chi^2= 3.84$.
We checked explicitly for a number of representative DM masses that constructing the profile likelihood from a separate scan with fixed mass
gives limits in good agreement with the above procedure.

Compared to the previous analysis \cite{Cuoco:2017rxb} we improve the coverage of the final \textsc{MultiNest} likelihood points sample 
before calculating the limits. We merge two different strategies. 
In the first, we exploit the fact that the $\sv$ parameter enters linearly in the propagation equation
and thus the likelihood at different values of $\sv$, and for fixed values of the remaining parameters,
can be  calculated using a single \textsc{Galprop} output and simply rescaling the DM contribution.
We thus take the \textsc{Galprop} output for the points collected in the scan, and we
use it, for each point, to derive the likelihood on a grid of 41 point 
logarithmically spaced in $\sv$ from  $\sv= 10^{-27}$cm$^3$s$^{-1}$ to $\sv=10^{-23}$cm$^3$s$^{-1}$.
In this way, we effectively increase the number of sample points by a factor of  40. 
Secondly, we merge together sample points collected from different scans.
For example, we extract relevant points of the $b\bar{b}$ channel, change the final state to $gg$, rerun \textsc{Galprop}, 
reevaluate the likelihood, and finally add these points to the sample of $gg$ points. 
Combined, the two strategies provide a better coverage which exploits the degeneracy between DM and propagation parameters in the fit and result in more robust limits.

This new approach allows to significantly reduce statistical fluctuations in the derived limits,
and to precisely compare the results for the various final states. We find that the limit of 
$WW/ZZ$ and $gg/c \bar c$ are basically indistinguishable. Furthermore, the limits of the leptonic 
channels only differ by a constant factor of $\nu/\ell \simeq 0.4$. 

%------------------------------------------------------------------------
\subsection{Limits from $\gamma$-ray observations of dwarf spheroidal galaxies}
%------------------------------------------------------------------------

Dark matter annihilation can also be 
tested by $\gamma$-ray observations of dwarf spheroidal galaxies.
We derive the corresponding limits for all annihilation channels considered here 
from recent Fermi-LAT data~\cite{Fermi-LAT:2016uux}. We use the published likelihoods 
provided for individual dwarfs as a function of the energy flux in the considered 24 energy bins. 

The predicted energy flux in an energy bin between $E_{\min}$ and $E_{\max}$ is given by
\begin{equation}
E^2 \frac{\D \phi}{\D E}=\frac{1}{4\pi} \frac{\langle\sigma v\rangle}{2 m_\text{DM}^2} \int_{E_{\min}}^{E_{\max}} \D E_\gamma \,E_\gamma\frac{\D N_\gamma}{\D E_\gamma}\times \int_\text{ROI} \D \Omega \int_\text{l.o.s}\D s \, \rho_\text{DM}^2  \,,
\end{equation}
where $\D N_\gamma/\D E_\gamma$ is the differential photon spectrum per annihilation,
 and $\rho_\text{DM}$ is the DM density
profile of the given dwarf. The integral over the region of interest (ROI) and line of sight (l.o.s) is called $J$-factor. 
We use the prediction for the differential photon spectrum 
from~\cite{Cirelli:2010xx} which includes electroweak corrections.

The total log-likelihood is obtained by summing over the log-likelihood contributions of the
nine dwarfs with the largest confirmed $J$-factors as given in  \cite{Geringer-Sameth:2014yza}.
We marginalize over the $J$-factor for each dwarf according to its uncertainty
given in~\cite{Geringer-Sameth:2014yza}.

%------------------------------------------------------------------------
\subsection{Limits from $\gamma$-ray observations towards the Galactic center}\label{sec:HESS}
%------------------------------------------------------------------------

\begin{table}[t]
\begin{center}
\renewcommand{\arraystretch}{1.2}
\begin{tabular}{c | c } 
density profile  & $\log_{10}\left( J_\text{HESS}/(\text{GeV}^2\,\text{cm}^{-5})\right)$ \\ \hline
Einasto & 21.66\\
NFW & 21.41\\
Burkert ($r_\text{c}=5\,$kpc) & 20.05\\
Burkert ($r_\text{c}=10\,$kpc) & 19.34\\ \hline 
\end{tabular}
\renewcommand{\arraystretch}{1}
\end{center}
\caption{$J$-factors relevant for H.E.S.S.\ $\gamma$-ray searches in the central Galactic halo region.}
\label{tab:JHESS}
\end{table}

For heavy DM, in particular, H.E.S.S. has provided bounds from $\gamma$-ray observations towards the central Galactic halo region~\cite{Abdallah:2016ygi}. However, these bounds depend sensitively on the DM profile close to the center of the Galaxy, which is not well constrained by observations, see also~\cite{Pierre:2014tra}. In the region relevant for the H.E.S.S. limits (a circle of $1^{\circ}$ radius centered on the Galactic center, with the Galactic plane excluded by requiring the latitude $|b| > 0.3^{\circ}$) the ratio of the $J$-factors between the NFW and Burkert $r_\text{c}=5 \,(10)$\,kpc profiles adopted in our cosmic-ray analysis  is approximately a factor of ~20\,(100).\footnote{The Burkert profile with a core radius of $r_c=10\,$kpc has been favored by a recent observational analysis~\cite{Nesti:2013uwa}.}
For further comparison we also adopt the Einasto profile~\cite{Einasto:1965czb}, 
$\rho_{\mathrm{Ein}}(r) =  \rho_\text{h}\exp\!\left( - 2/\alpha \left[(r/r_\text{h})^2-1\right]\right)$, with a characteristic halo radius $r_\text{h}=20\,$kpc, and a characteristic halo density $\rho_\text{h}$, 
normalized so that to obtain a local DM density $\rho_\odot = 0.43\,$GeV/cm$^3$~\cite{Salucci:2010qr} 
at the solar position, $r_ \odot = 8\,$kpc. 
In table~\ref{tab:JHESS} we collect the $J$-factors relevant for the interpretation of the H.E.S.S. limits for all considered benchmark DM profiles. Note, however, that the analysis of~\cite{Abdallah:2016ygi} has been optimized for cusped profiles such as NFW or Einasto. Rescaling the results presented in \cite{Abdallah:2016ygi} with the $J$-factors of cored profiles is thus only an approximate way to derive the corresponding dark matter limits.
We will use the $J$-factors of table~\ref{tab:JHESS}  also in Section~\ref{sec:models}  when we discuss the H.E.S.S.
constraints on $\gamma$-ray lines from the Galactic halo region \cite{Abramowski:2013ax}. 

%------------------------------------------------------------------------
\subsection{Results and discussion}\label{sec:resdis}
%------------------------------------------------------------------------

%=====================
%    \                                           |
%      \                                         |
%        \                                       |
\begin{figure}[tp]
\vspace{-0.5cm}
\centering
\setlength{\unitlength}{0.98\textwidth}
\begin{picture}(1,1.233)
  \put(0.0,0.94){\includegraphics[width=0.5\textwidth]{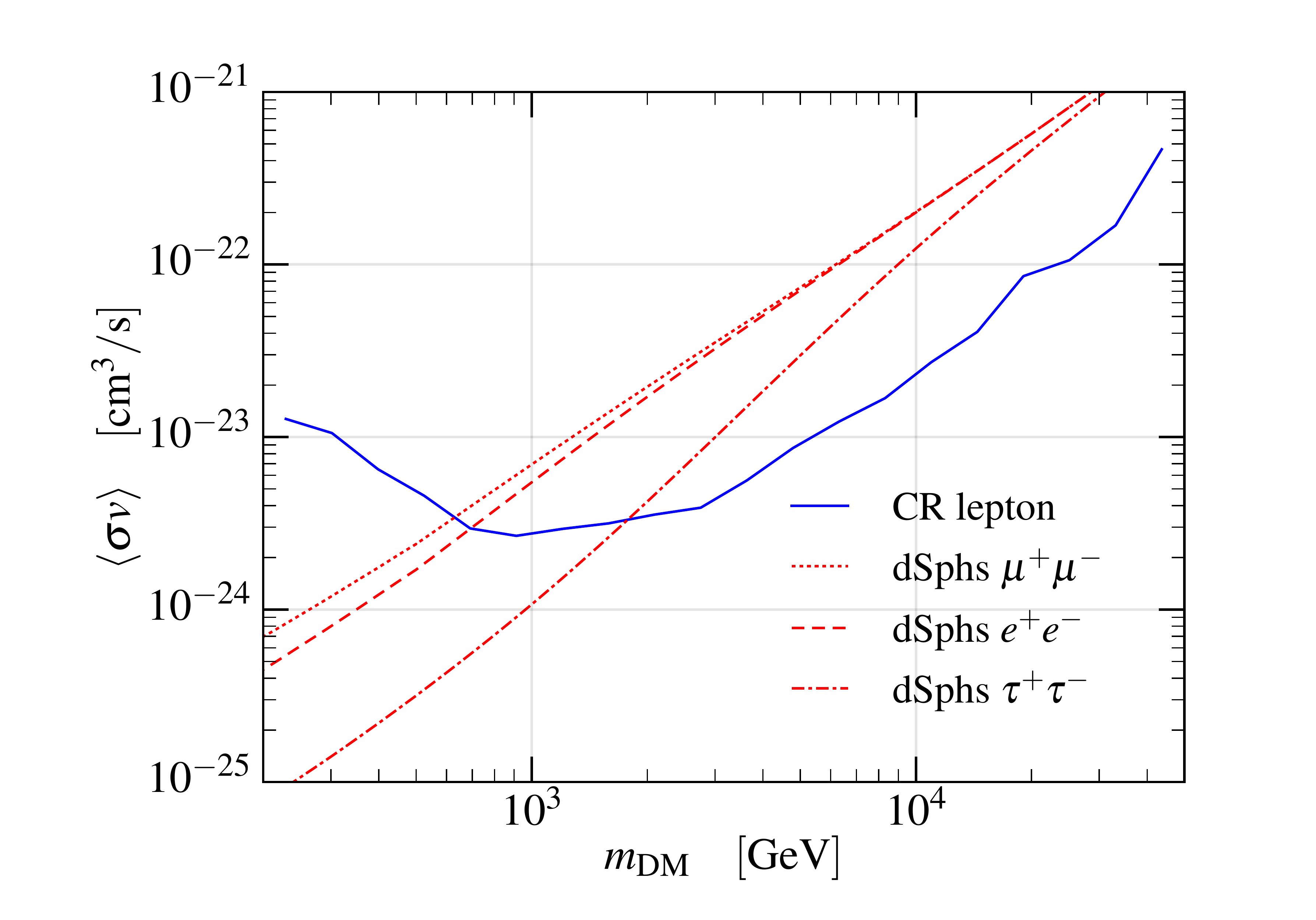}}
  \put(0.495,0.94){\includegraphics[width=0.50\textwidth]{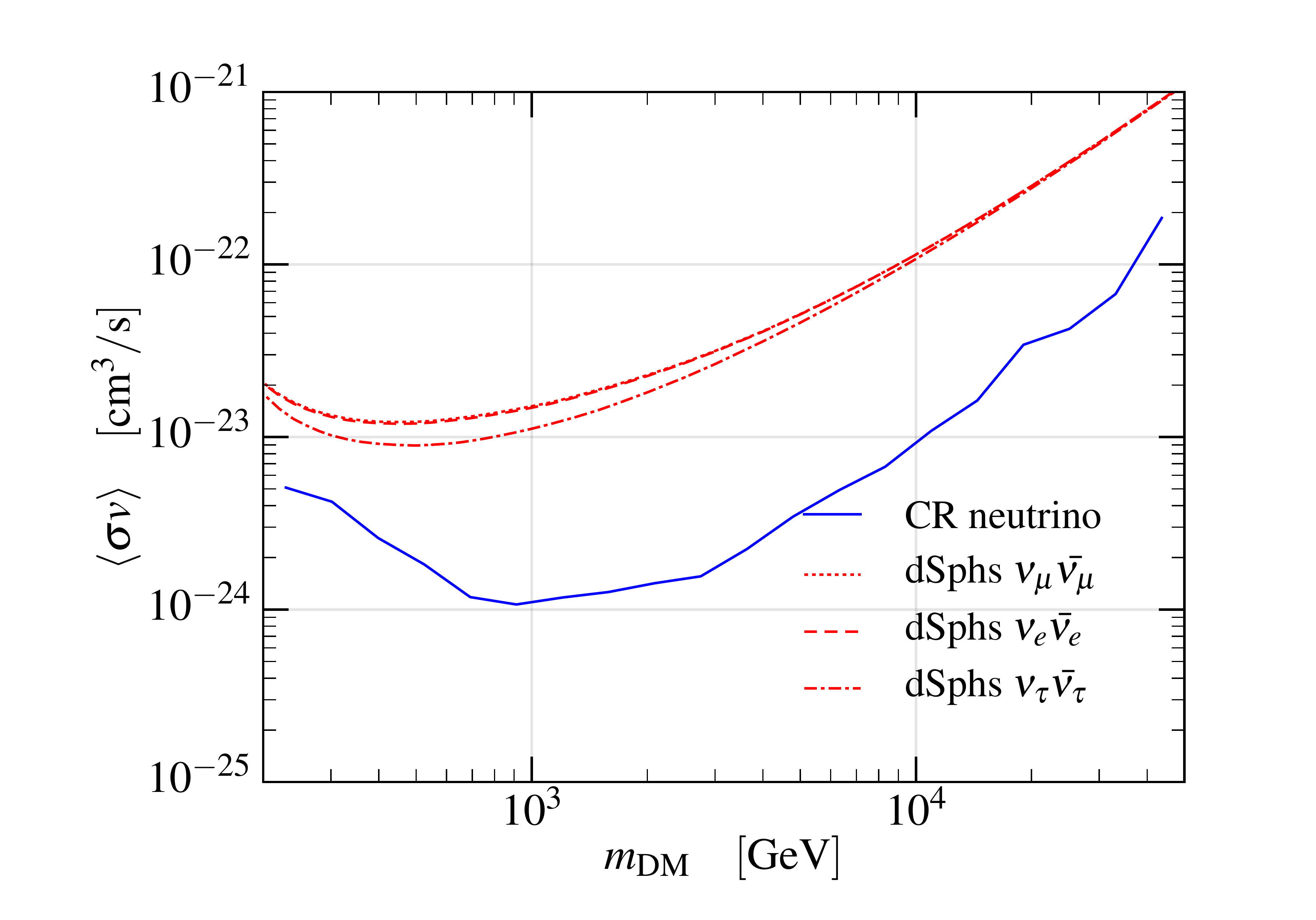}}
  \put(0.0,0.6){\includegraphics[width=0.50\textwidth]{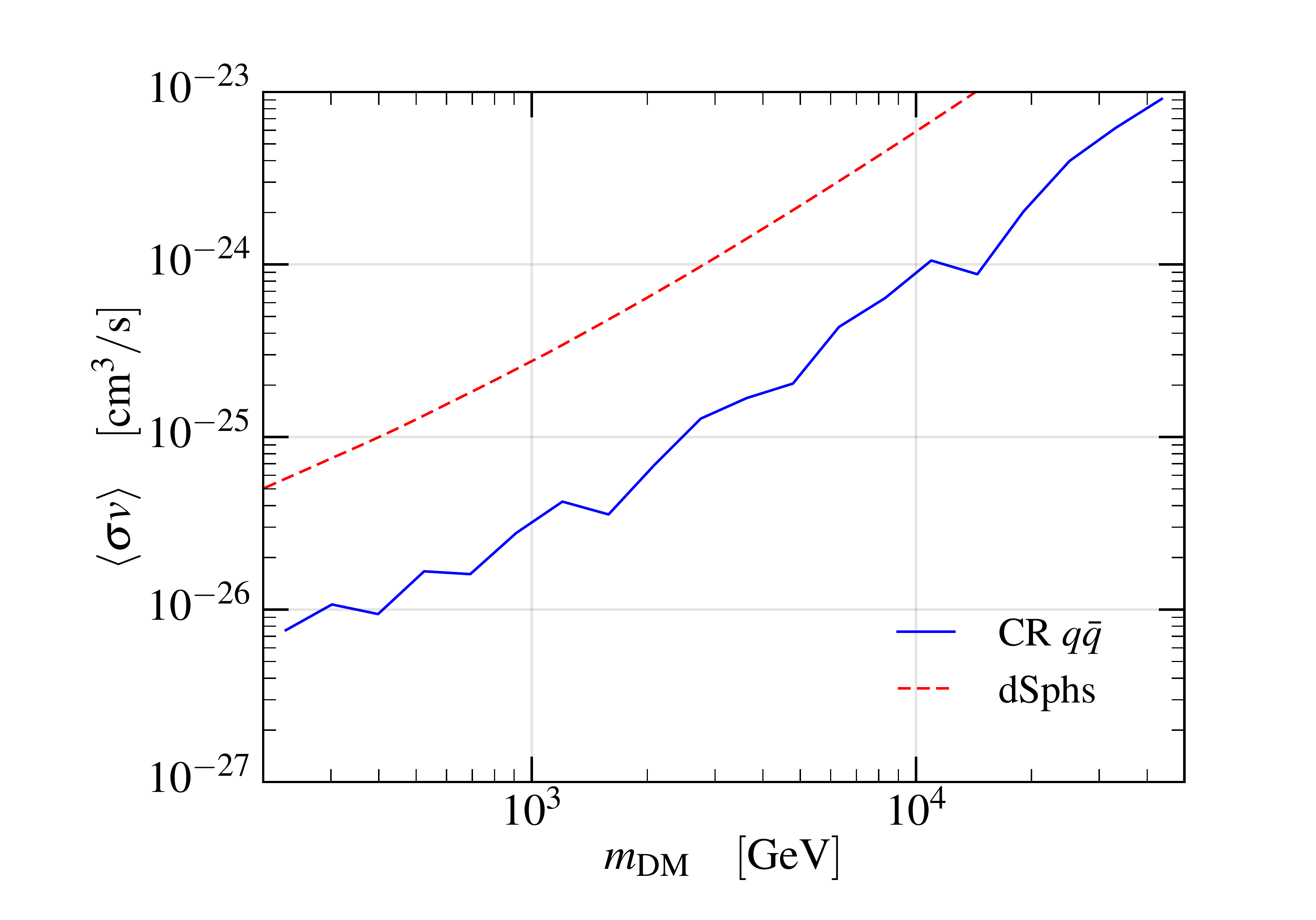}}
  \put(0.495,0.6){\includegraphics[width=0.50\textwidth]{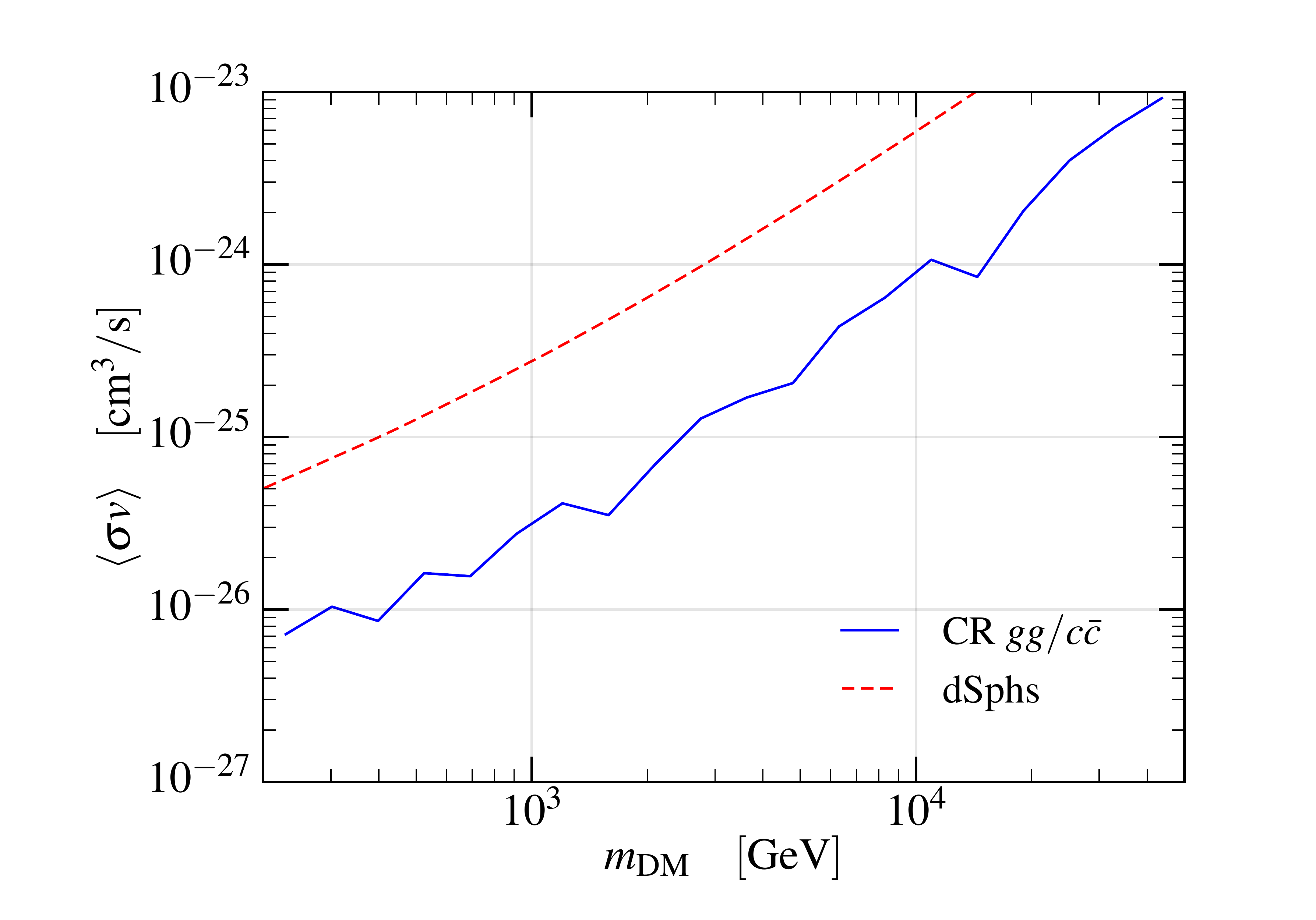}}
  \put(0.0,0.26){\includegraphics[width=0.50\textwidth]{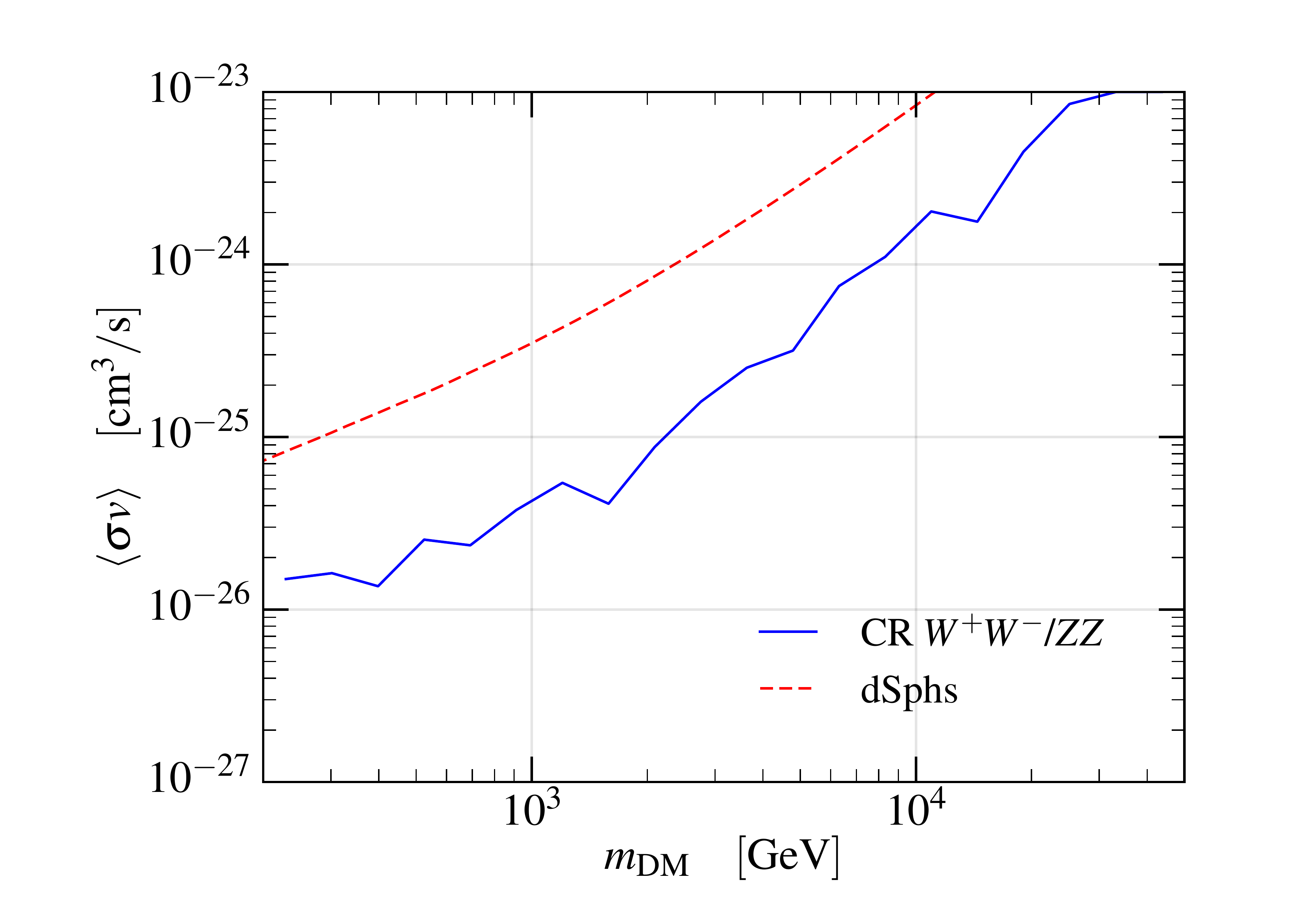}}
  \put(0.495,0.26){\includegraphics[width=0.50\textwidth]{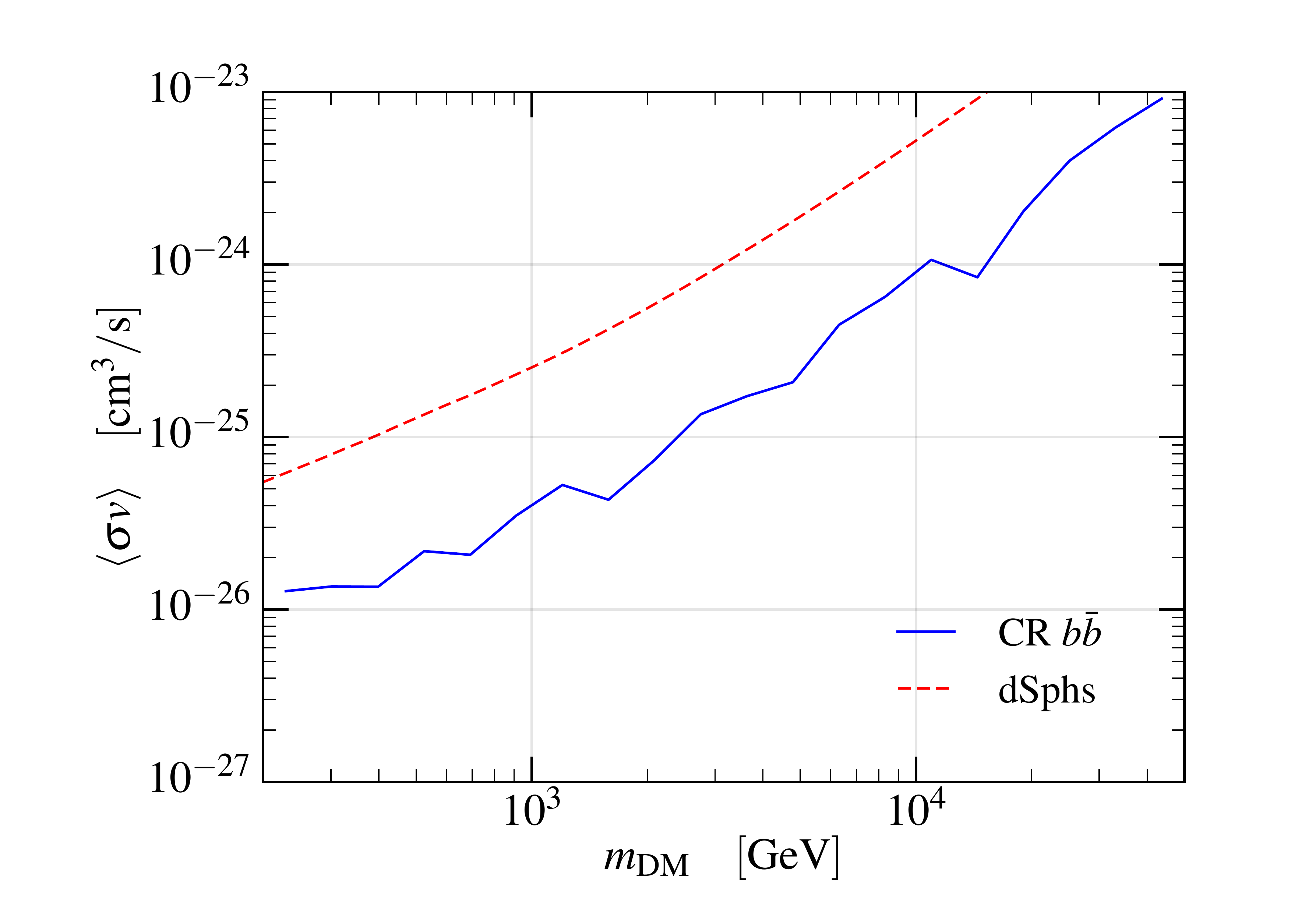}}
  \put(0.0,-0.08){\includegraphics[width=0.50\textwidth]{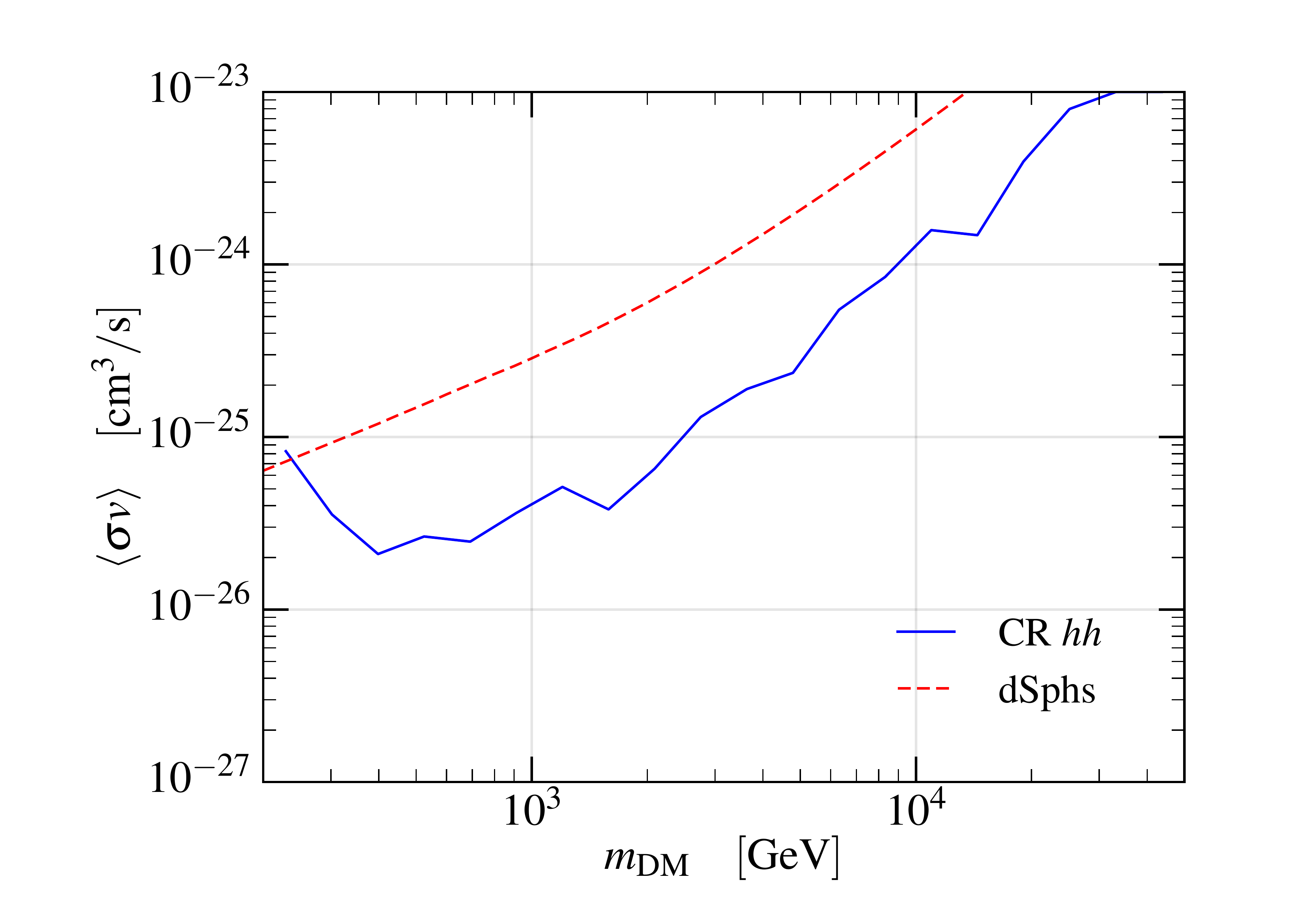}}
  \put(0.495,-0.08){\includegraphics[width=0.50\textwidth]{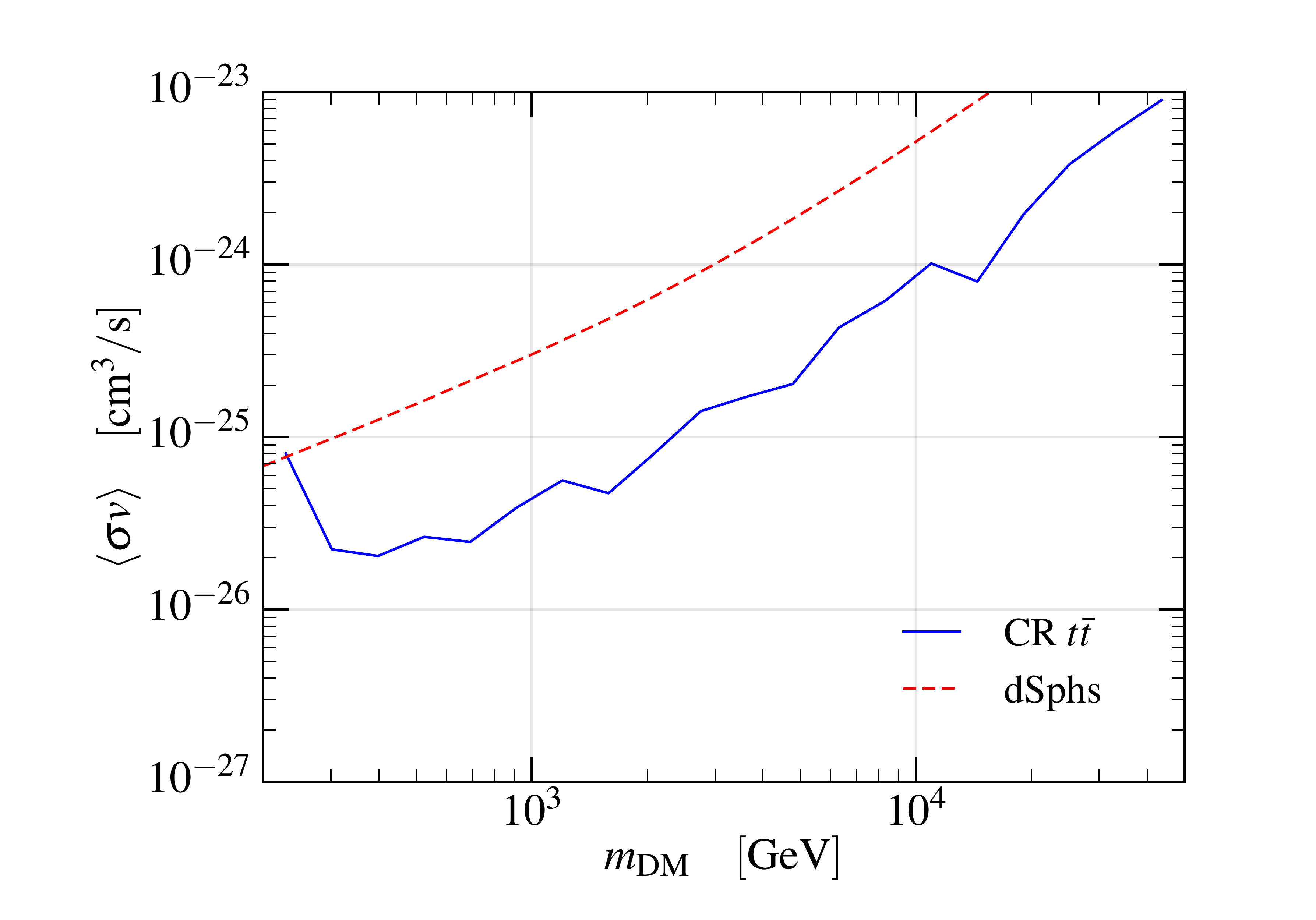}}
\end{picture}
\vspace{1cm}
\caption{95\% C.L. upper limits on the DM annihilation cross section $\sv$ for all possible annihilation channels into pairs of
SM particles from CR antiprotons (solid blue curves) and from dwarf spheroidal 
galaxies (dashed red curves). For the leptonic channels the CR limits are flavor
blind. Note the different scales on the vertical axes when comparing the limits of the leptonic (upper row) and non-leptonic annihilation channels.}
\label{fig:jointind}
\end{figure}
%                                      \         |
%                                        \       |
%                                          \     |
%=====================

The 95\% C.L. limits on the DM annihilation cross section $\sv$ from our analysis of the AMS-02 
antiproton spectra are shown in figure~\ref{fig:jointind} for DM masses between 200~GeV and 50~TeV. 
We present results for all SM final states, $f=q, c, b, t, g, W,Z, h, \ell$ and $\nu$. Note that the limits 
for $W/Z$ and for $g/c$ final states are virtually identical and are thus not displayed separately. Furthermore, the limits for the leptonic 
channels are flavour-independent, and differ between charged leptons and neutrinos by a constant factor of $\nu/\ell \simeq 0.4$.

For comparison, we also show the limits derived from the Fermi-LAT observations of dwarf spheroidal galaxies. The limits from $\gamma$-ray observations towards the Galactic center depend sensitively on the DM profile and will thus be discussed separately below. For DM annihilation into quarks or 
gauge bosons, the antiproton limits are significantly stronger than those derived from dwarf galaxies. 
For DM annihilating into Higgs particles or top quarks, the antiproton limits become comparable to those from dwarf galaxies for DM masses below about 300~GeV, i.e.\ in the mass region 
consistent with a potential DM signal in the antiproton spectrum~\cite{Cuoco:2017rxb}. 

Even for the leptonic channels, where antiprotons only arise from electroweak 
radiation, we find competitive constraints from our analysis of the AMS-02 antiproton data, in particular for large DM masses where electroweak radiation is logarithmically enhanced, see also~\cite{Cavasonza:2016qem}. Note that the limits on annihilation 
into leptonic finals states are in general considerably weaker then those on annihilation into gluons, quarks, gauge or Higgs bosons.

\begin{figure}[t]
\vspace{0.5cm}
\centering
\setlength{\unitlength}{1\textwidth}
\begin{picture}(1,0.6)
  \put(0.0,0.3){\includegraphics[width=0.478\textwidth]{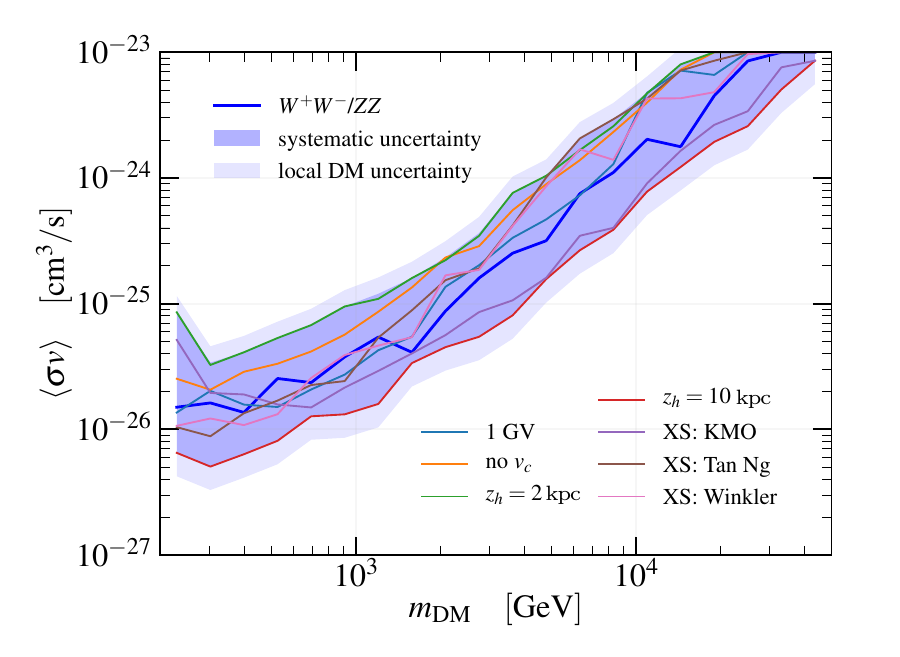}}
  \put(0.495,0.3){\includegraphics[width=0.51\textwidth]{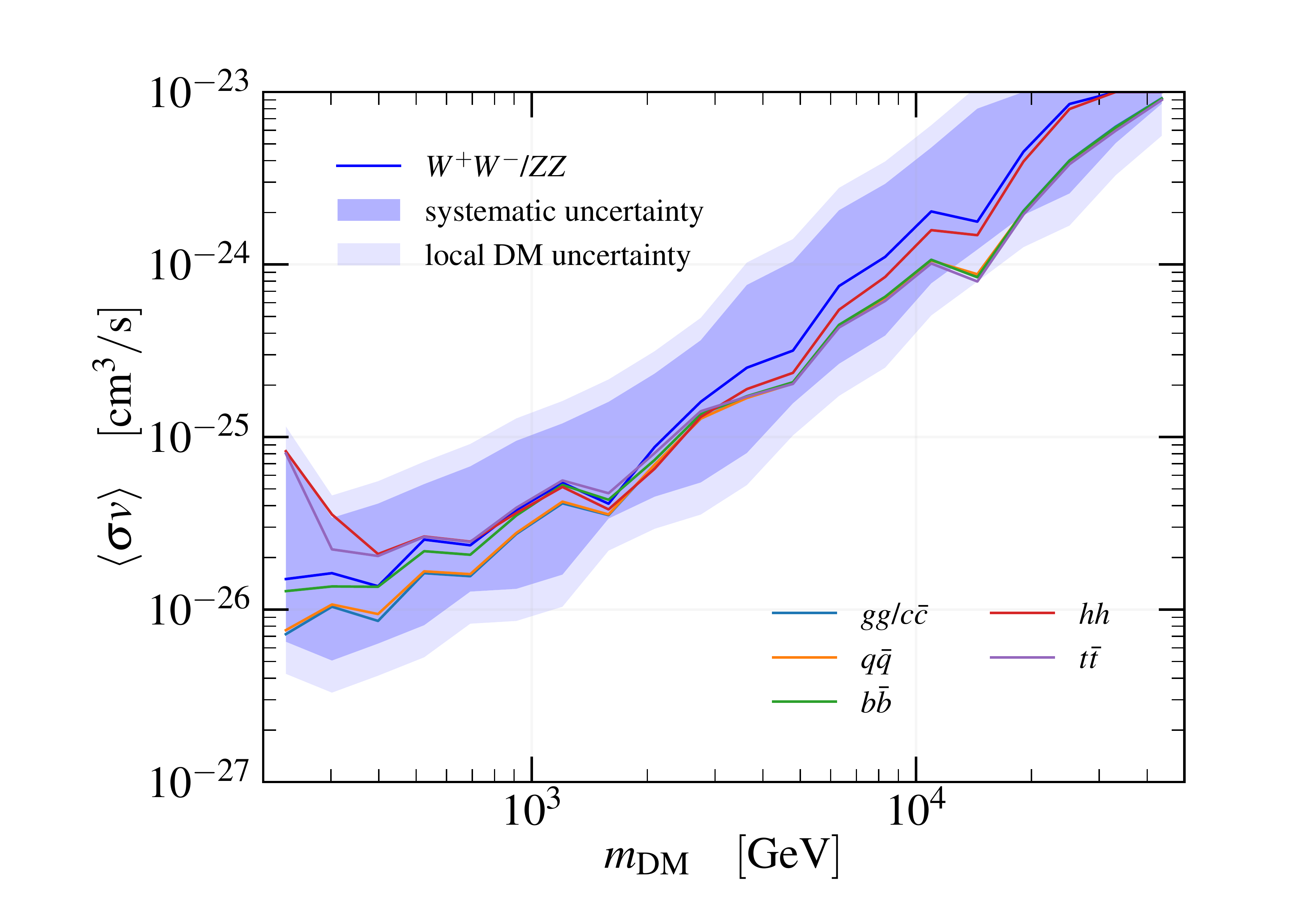}}
  \put(0.245,-0.05){\includegraphics[width=0.51\textwidth]{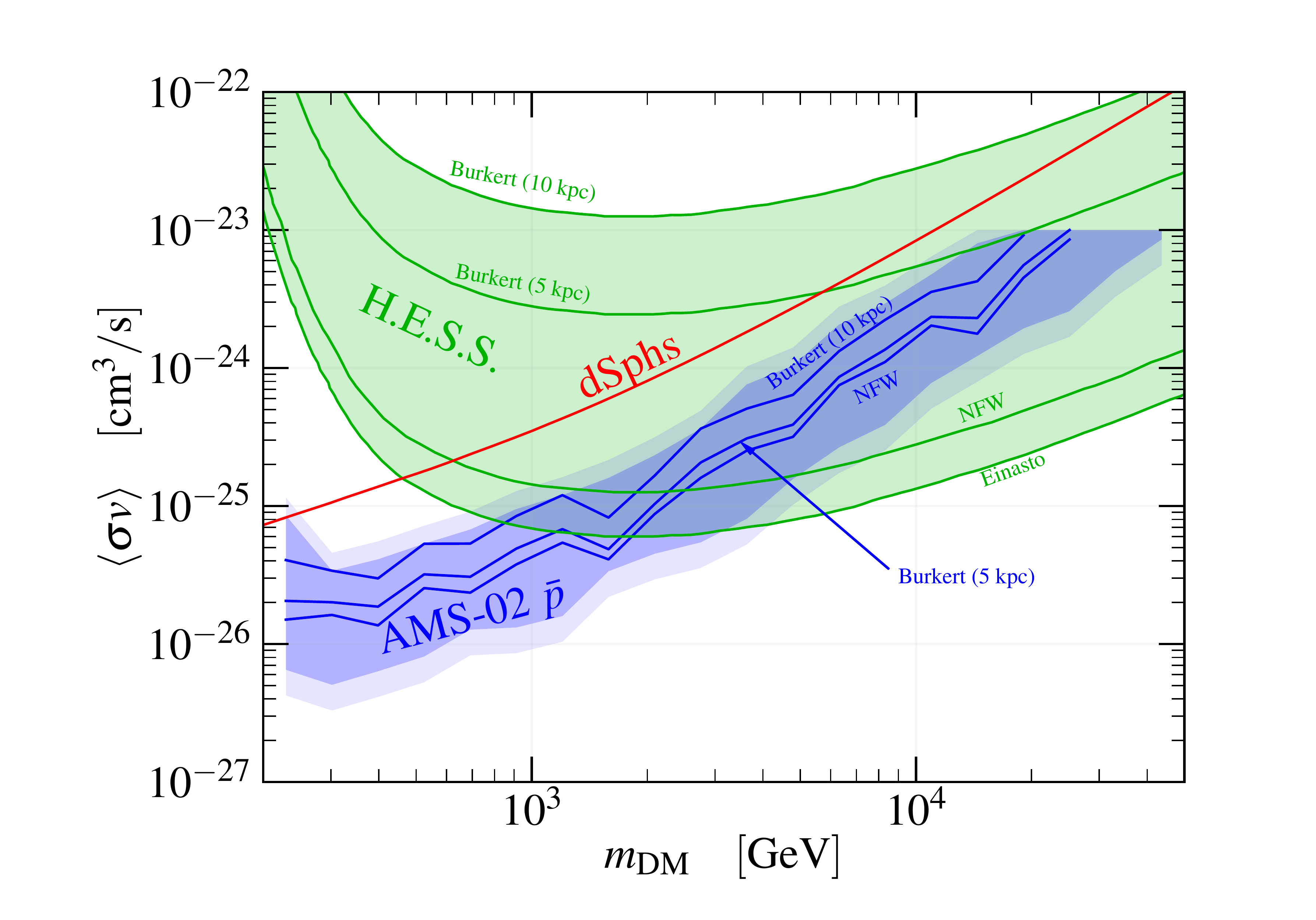}}
\end{picture}
\caption{95\% C.L. upper limits on the DM annihilation cross section for $WW/ZZ$ final states. Upper left panel: Antiproton limits for a variety of propagation settings
and antiproton cross-section predictions. The envelope of all curves determines the dark blue shaded band and indicates 
the overall systematic uncertainty. The light blue shaded band denotes the additional uncertainty from the
local DM density. Upper right panel: CR antiproton limits on the DM annihilation cross section for non-leptonic annihilation channels, compared to the limits for $WW/ZZ$ final states including the overall systematic uncertainty.
Lower panel: Comparison of CR antiproton and $\gamma$-ray limits, including the uncertainty from the Galactic DM density profile.}
\label{fig:uncert}
\end{figure}

It is important to quantify the uncertainty on the antiproton limits arising from our description of cosmic ray propagation, the antiproton cross section, as well as the DM density profile. We have thus 
repeated our analysis with a diffusion model without convection and with a fixed Galaxy half-height of $z_h = 10\,$kpc and $2\,$kpc, respectively. We have also studied the impact of extending the range 
of rigidities included in our fit down to 1\,GV (see the discussion in \cite{Korsmeier:2016kha}). 
To estimate the uncertainty introduced by the antiproton cross section, we  
adopt different parametrizations as provided in Refs.~\cite{Mauro_Antiproton_Cross_Section_2014, Donato:2017ywo, TanNg_AntiprotonParametrization_1983, Kachelriess:2015wpa,Kappl:2014hha,Winkler:2017xor}. 
The DM limits corresponding to the different setups are displayed in figure~\ref{fig:uncert} (upper left panel) for annihilation into $WW/ZZ$, which is the final state relevant for the specific DM models studied in section~\ref{sec:models}. 
As the overall systematic uncertainty on the cross-section limit we consider the envelope of the various limits, as indicated by the dark blue shaded region in  figure~\ref{fig:uncert}. In addition, we display the uncertainty from the local DM density, $\rho_\odot = (0.43 \pm 0.15)\,$GeV/cm$^3$, which we add linearly to the other systematic uncertainties (light blue shaded region). The uncertainty introduced by the DM density profile falls within the uncertainty band from CR propagation and the antiproton cross section, and will be quantified below. 

In figure~\ref{fig:uncert}, upper right panel, we compare the CR antiproton limits obtained for the annihilation into $WW/ZZ$ with those into gluons, quarks or Higgs particles. The limits are quite similar above DM masses of about 500~GeV, in particular in comparison with the systematic uncertainty indicated by the blue shaded region. Thus, it is a good and conservative approximation to use the DM annihilation cross-section limits derived for $WW/ZZ$ final states for probing DM models with annihilation into gluons, quarks or Higgs bosons, as long as the DM mass is above approximately 500~GeV. 

Finally, in figure~\ref{fig:uncert}, lower panel, we compare our CR antiproton limits on the DM annihilation cross section for $WW/ZZ$ final states with the limits obtained by H.E.S.S.\  from $\gamma$-ray observations towards the Galactic center. 
To quantify the dependence of the antiproton limits on the DM density profile we have compared the results for the default NFW profile with those obtained from a  Burkert profile with core radii of $r_c=5\,$kpc and $r_c=10\,$kpc, respectively. 
The H.E.S.S.\ limits for the NFW and Einasto profiles are taken from~\cite{Abdallah:2016ygi}, the limits for the Burkert profiles have been obtained in an approximate way by rescaling with the corresponding $J$-factors of table~\ref{tab:JHESS}, see section~\ref{sec:HESS}. For completeness, we also show the limit derived from the Fermi-LAT observations of dwarf spheroidal galaxies. While the H.E.S.S.\ limits for the NFW or Einasto profiles are stronger than those from CR antiprotons for DM masses beyond ${\cal O}$(TeV), they are subject to very large uncertainties from the DM profile near the central Galactic halo region and become significantly less competitive for cored profiles.

%=====================
%    \                                           |
%      \                                         |
%        \                                       |
%                                      \         |
%                                        \       |
%                                          \     |
%=====================

%===================================================================
\section{Constraints on minimal dark matter models}\label{sec:models}
%===================================================================

In this section we derive constraints on minimal DM models from cosmic ray antiproton fluxes and compare these to other constraints from indirect detection. We consider SM extensions with an electroweak multiplet fermion, $\chi$, 
where
\begin{equation}
{\cal L} = {\cal L}_{\rm SM} + \bar\chi (iD\!\!\!\!\slash +M)\chi\,. 
\end{equation}
The interactions of the fermion multiplet with the SM are determined by the covariant derivative, $D_\mu$. The only free parameter of such models is the tree-level mass of the fermion multiplet, $M$. Radiative corrections induce a mass splitting such that the lightest component of $\chi$ is neutral and thus a DM candidate. 
The requirement of a neutral DM candidate restricts the choice of the hypercharge $Y$ of the electroweak multiplet~\cite{Cirelli:2005uq}.

We focus on three particularly interesting scenarios, where $\chi$ is a fermion doublet, triplet or quintuplet. 
The minimal models with a fermion doublet (with $Y=1/2$) and triplet (with $Y=0$) correspond to well-motivated limits of supersymmetric theories, i.e.\  higgs\-ino or wino DM, respectively. The quintuplet, on the other hand, is the simplest representation where the DM candidate is stable without imposing an additional symmetry beyond gauge symmetry and Poincar\'{e} invariance. These models can provide the correct relic density for DM masses in the multi-TeV region, see below.  

The currently most powerful way to probe minimal DM with electroweak fermions is through indirect detection. 
In our minimal models, DM annihilates predominantly into SM gauge bosons, i.e.\ $\chi\chi \to W^+W^-, ZZ$, $Z\gamma$ and $\gamma\gamma$. As discussed in section~\ref{sec:limits}, the decay and fragmentation of the $W$ and $Z$ bosons produce all types of SM particles, leading to a continuous spectrum of photons (mostly from pion decay) as well as positrons and antiprotons. Furthermore, the annihilation channels $Z\gamma$ and $\gamma\gamma$ give rise to monochromatic photons providing spectral lines. The annihilation cross section is determined by the SM gauge couplings and the DM mass. For heavy electroweak DM, the annihilation cross section is strongly enhanced by the Sommerfeld effect, see e.g.~\cite{Hisano:2003ec,Hisano:2004ds,Hisano:2006nn,Hryczuk:2010zi,Hryczuk:2011vi,Chun:2012yt,Hryczuk:2014hpa,Baumgart:2014vma,Bauer:2014ula,Ovanesyan:2014fwa,Baumgart:2014saa,Baumgart:2015bpa,Chun:2015mka,Ovanesyan:2016vkk,Beneke:2016jpw}, resulting in potentially large indirect detection signals. 

In this section we constrain the model parameter space of minimal DM by limits from antiprotons, continuous $\gamma$ rays and $\gamma$ lines. Limits from positrons~\cite{Kopp:2013eka} and CMB constraints~\cite{Kawasaki:2015peu} are significantly weaker than those from antiprotons and photons. The remainder of the section 
is structured as follows: In sections~\ref{sec:wino}, \ref{sec:higgsino} and \ref{sec:5plet} we briefly review phenomenological aspects of the wino, higgsino and quintuplet models, respectively. 
Finally, in section~\ref{sec:MDMresults} we present and discuss our results.

%------------------------------------------------------------------------
\subsection{Wino dark matter} \label{sec:wino}
%------------------------------------------------------------------------

An $SU(2)_L$ triplet fermion multiplet with zero hypercharge is predicted in the minimal supersymmetric extension of the Standard Model (MSSM). Such a wino multiplet consists of a neutral Majorana fermion, $\chi^0$, and a charged fermion, $\chi^\pm$, the superpartners of the $SU(2)_L$ gauge bosons. Loop effects generate a mass splitting between the neutral and the charged states of around 160\,MeV~\cite{Cheng:1998hc,Feng:1999fu,Gherghetta:1999sw,Yamada:2009ve,Ibe:2012sx}.
In theories with anomaly-mediated supersymmetry breaking~\cite{Bagnaschi:2016xfg}, the neutral wino is the lightest superparticle and thus a natural DM candidate. Wino DM has been studied in great detail in the literature, see e.g.~\cite{Cirelli:2007xd,Chun:2012yt,Cohen:2013ama,Fan:2013faa,Bhattacherjee:2014dya,Chun:2015mka}. We assume a minimal scenario, where all superparticles except for the wino multiplet are decoupled; for a more general analysis of wino-like DM in supersymmetric models see e.g.~\cite{Beneke:2016ync,Beneke:2016jpw}. 

A thermal wino with mass $m_\chi \approx 2.8$~TeV provides the correct relic density~\cite{Hisano:2006nn,Cirelli:2007xd,Hryczuk:2010zi}. 
While the LHC is not sensitive to minimal models with TeV-scale wino DM~\cite{Chen:1996ap,Asai:2008sk,Low:2014cba,Cirelli:2014dsa}, a 100~TeV collider should be able to probe thermal wino scenarios through searches for disappearing charged tracks from $\chi^\pm \to \chi^0\pi^\pm$ decays~\cite{Low:2014cba,Cirelli:2014dsa,Bramante:2015una,Mahbubani:2017gjh}. 
Current and future LHC searches are sensitive to light wino DM with a mass around a few hundred GeV, which would however require a non-thermal history.  
The current experimental lower bound on the mass of pure wino DM, corresponding to a mass splitting of 160\,MeV
between the neutral and the charged states, is around 270\,GeV arising from searches for disappearing tracks
at the 8\,TeV LHC~\cite{Aad:2013yna} (see also~\cite{CMS:2014gxa}). A preliminary 
analysis of the 13\,TeV LHC run using $36.1\,\text{fb}^{-1}$ of data excludes masses up to 430\,GeV~\cite{ATLAS:2017bna}.

As wino DM has hypercharge zero and thus vanishing interactions with nucleons at tree-level, the direct detection signal~\cite{Cirelli:2005uq,Hisano:2010fy,Hisano:2010ct,Cirelli:2014dsa,Nagata:2014wma,Hill:2013hoa,Hill:2014yka,Hill:2014yxa,Hisano:2015rsa} 
is well below current bounds~\cite{Akerib:2016vxi,Aprile:2017iyp}. A multi-ton target direct detection experiment such as DARWIN~\cite{Aalbers:2016jon} would be required to discover or exclude minimal wino DM.

We probe wino DM by comparing the predicted annihilation cross section, $\sv$, with the bounds derived from indirect detection. 
As mentioned above, an accurate prediction of $\sv$ requires the inclusion of higher-order electroweak effects. We adopt the calculation of ~\cite{Hryczuk:2011vi}, performed at 1-loop level and including the electroweak Sommerfeld effect.\footnote{Calculations of DM annihilations into $\gamma$ or $Z$ final states in an effective field theory framework have been presented in~\cite{Baumgart:2014vma,Bauer:2014ula,Ovanesyan:2014fwa,Baumgart:2014saa,Baumgart:2015bpa,Ovanesyan:2016vkk}.} Tree-level annihilation into $WW$ is dominant (at around 90\%), as annihilation into $ZZ$, $Z\gamma$ and $\gamma\gamma$ is induced at loop level only.  
For an estimate of the uncertainties arising from the predictions of the shape of the antiproton spectrum, we confront our default choice~\cite{Cirelli:2010xx} with the spectra from~\cite{Hryczuk:2014hpa}, which include 1-loop and Sommerfeld corrections, see section~\ref{sec:MDMresults} for further details. Finally, we use the relic density prediction from~\cite{Cirelli:2007xd}.

%------------------------------------------------------------------------
\subsection{Higgsino dark matter}\label{sec:higgsino}
%------------------------------------------------------------------------

In the minimal DM model with a  fermion doublet with hypercharge $Y=1/2$, the DM particle is a Dirac fermion and has a large spin-independent elastic nucleon cross sections mediated by $Z$-boson exchange. Such models are thus excluded by current direct detection bounds. 

However, these bounds can be evaded in extended models such as supersymmetry. Because of mixing with the superpartners of the SM gauge bosons, 
the fermion doublet (higgs\-ino) is a Majorana fermion with a vanishing nucleon scattering cross section at tree level. 
We thus consider the higgsino model as a limit of a more complex scenario, where the additional particles are well above the TeV-scale and thus do not affect the indirect detection signals on which we focus here. In such models, the higgsino-nucleon scattering cross section is well below the neutrino background of direct detection experiments~\cite{Hisano:2015rsa}. A more comprehensive analysis of the phenomenology of supersymmetric models with higgsino-like DM has been presented, e.g., in \cite{Bramante:2015una,Beneke:2016jpw}. 

In the limit where all new particles other than the higgsino are very heavy, the mass splitting between the neutral and charged
state of the doublet is around 340\,MeV~\cite{Cirelli:2007xd} resulting in a decay length of ${\cal O}(1\,{\rm cm})$ of the charged state.
From the non-observation of disappearing charged tracks at the 13\,TeV LHC~\cite{ATLAS:2017bna} (preliminary analysis) a lower limit on the DM mass of around 120\,GeV can be derived~\cite{Fukuda:2017jmk}, see also~\cite{Mahbubani:2017gjh,Ostdiek:2015aga}. A thermal higgsino with mass $m_\chi \approx 1050$\,GeV provides the correct relic density~\cite{Cirelli:2007xd}.
For the analysis below we use the cross section and relic density prediction from~\cite{Cirelli:2007xd}. For the higgsino
DM model, annihilation into $WW$ and $ZZ$ finals states are of approximately equal importance. 

%------------------------------------------------------------------------
\subsection{Fermion quintuplet dark matter}\label{sec:5plet}
%------------------------------------------------------------------------

As for the triplet, the quintuplet fermion has hypercharge zero and thus vanishing interactions with nucleons at tree-level. Loop-induced interactions generate (predominantly spin-in\-depen\-dent) interactions. The corresponding direct detection signal is, however, well below current bounds~\cite{Hisano:2015rsa}.

The quintuplet contains a neutral, a single and a double charged state. 
Searches for disappearing charged tracks at the 8\,TeV LHC~\cite{Aad:2013yna} provide a lower limit of around 290\,GeV on the DM mass~\cite{Ostdiek:2015aga}.
A thermal quintuplet with mass $m_\chi \approx 9.4$\,TeV provides the correct relic density~\cite{Cirelli:2015bda}.
We use the cross section and relic density predictions from~\cite{Cirelli:2015bda}.

%=====================
%    \                                           |
%      \                                         |
%        \                                       |
\begin{figure}[t]
\vspace{0.5cm}
\centering
\setlength{\unitlength}{1.0\textwidth}
\begin{picture}(1,0.57)
  \put(0.0,-0.038){\includegraphics[trim={40 40 40 40},clip, width=1\textwidth]{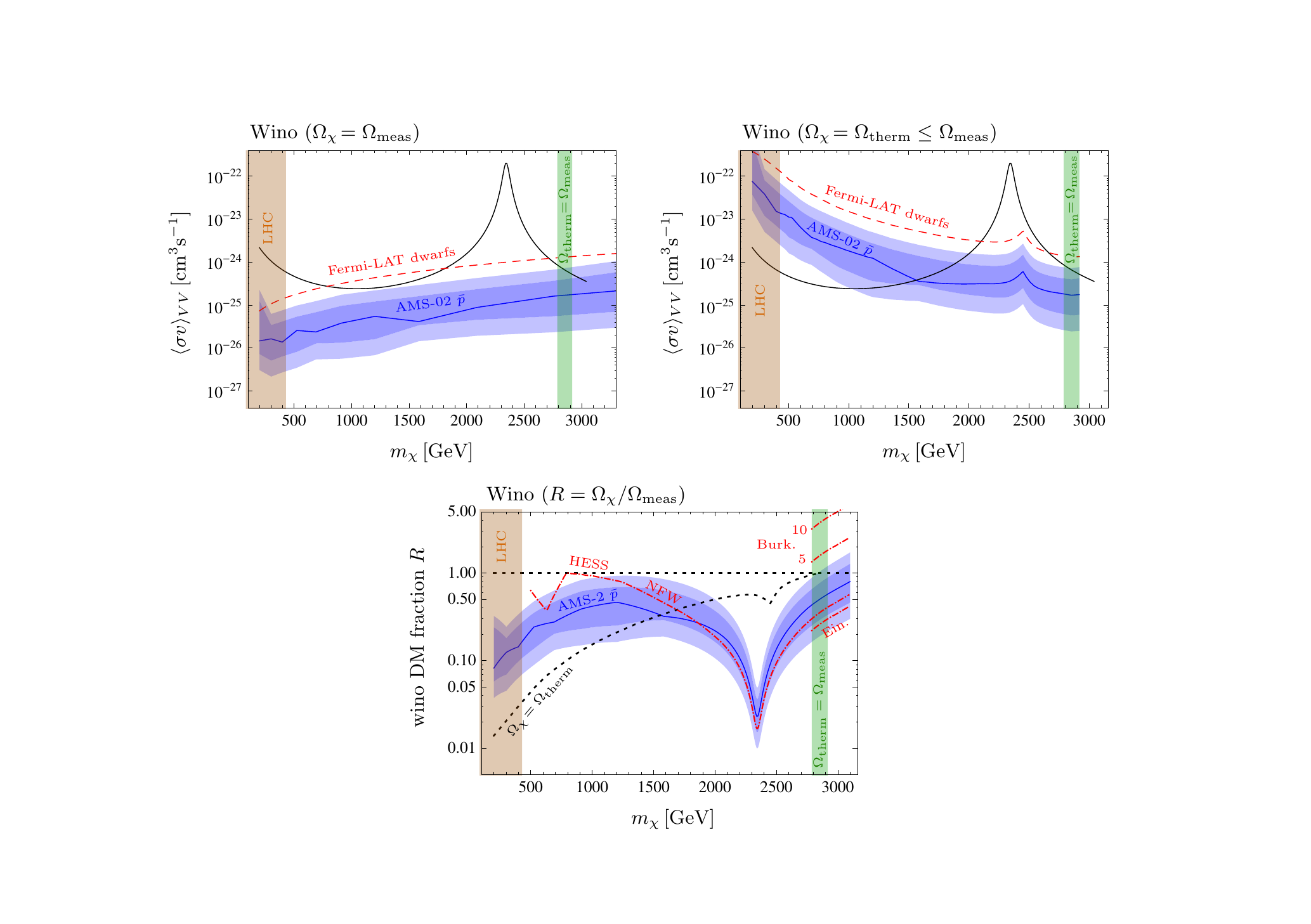}}
\end{picture}
\caption{95\% CL exclusion limit on minimal wino dark matter. The blue curve shows the 
upper limit from AMS-02 antiprotons. The dark and light blue shaded error bands display the
CR systematic uncertainties (see text for details) and the uncertainties from the variation 
of $\rho_\odot$ (added linearly), respectively. The vertical green shaded band around 2850\,GeV corresponds to the DM mass range 
where the thermal relic density matches the measured one. The brown shaded
band on the left denotes the mass range excluded by LHC searches. 
Upper panels: Limits on the annihilation cross section into vector bosons 
for two cases: 100\% wino DM (left panel) and a wino DM fraction according to the thermal
production (right panel). The red dashed curve shows the Fermi-Lat $\gamma$-ray limits from 
dwarf spheroidal galaxies. 
The solid black curves show the cross section prediction.
Lower panel: Upper limits in terms of the wino DM fraction $R$, i.e.\ the ratio of minimal DM to all of DM. For comparison 
we show the H.E.S.S.\ limits from searches for $\gamma$-lines from Galactic center observations
(red dot-dashed curves). The limit extending over the whole mass range above 500\,GeV
assumes the NFW profile. To reduce clutter the respective limits
for the Burkert 10\,kpc, Burkert 5\,kpc and Einasto profiles (from top to bottom) are only
displayed on the very right of the mass range. The relative difference between these choices is a constant factor.
The two black dotted curves illustrate the two cases considered in the upper panels, i.e.\ 100\% wino DM ($R=1$)
and $R$ according to the thermal production, $\Omega_\chi=\Omega_\text{therm}$. }
\label{fig:wino}
\end{figure}
%                                      \         |
%                                        \       |
%                                          \     |
%=====================

%------------------------------------------------------------------------
\subsection{Results and discussion}\label{sec:MDMresults}
%------------------------------------------------------------------------

\begin{figure}[t]
\vspace{0.5cm}
\centering
\setlength{\unitlength}{1.0\textwidth}
\begin{picture}(1,0.52)
  \put(0.0,-0.038){\includegraphics[trim={40 40 40 40},clip, width=1\textwidth]{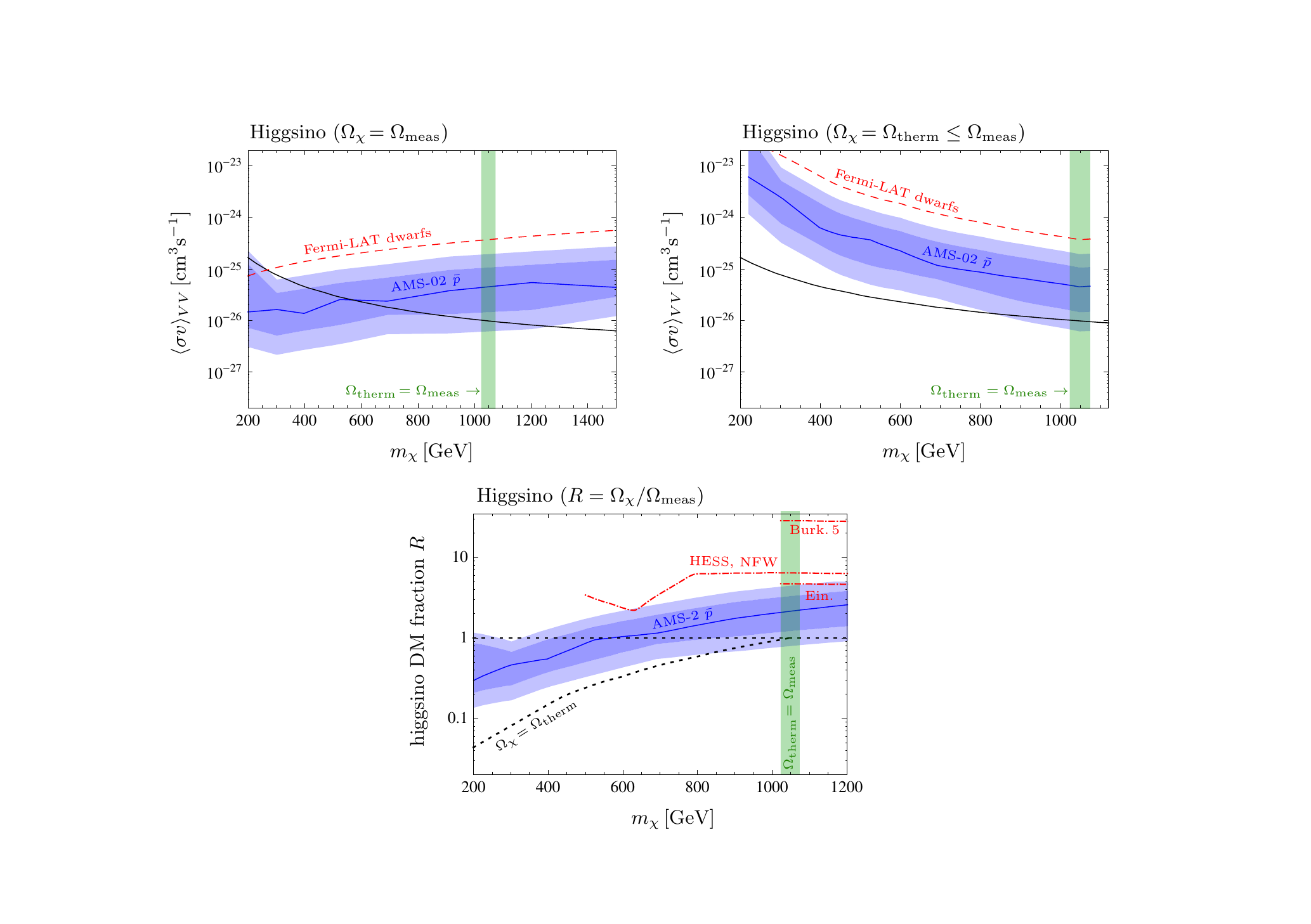}}
\end{picture}
\caption{Same as figure~\ref{fig:wino} but for higgsino dark matter.}
\label{fig:higgsino}
\end{figure}

\begin{figure}[t]
\vspace{1.9cm}
\centering
\setlength{\unitlength}{1.0\textwidth}
\begin{picture}(1,0.52)
  \put(0.0,-0.038){\includegraphics[trim={40 40 40 40},clip, width=1\textwidth]{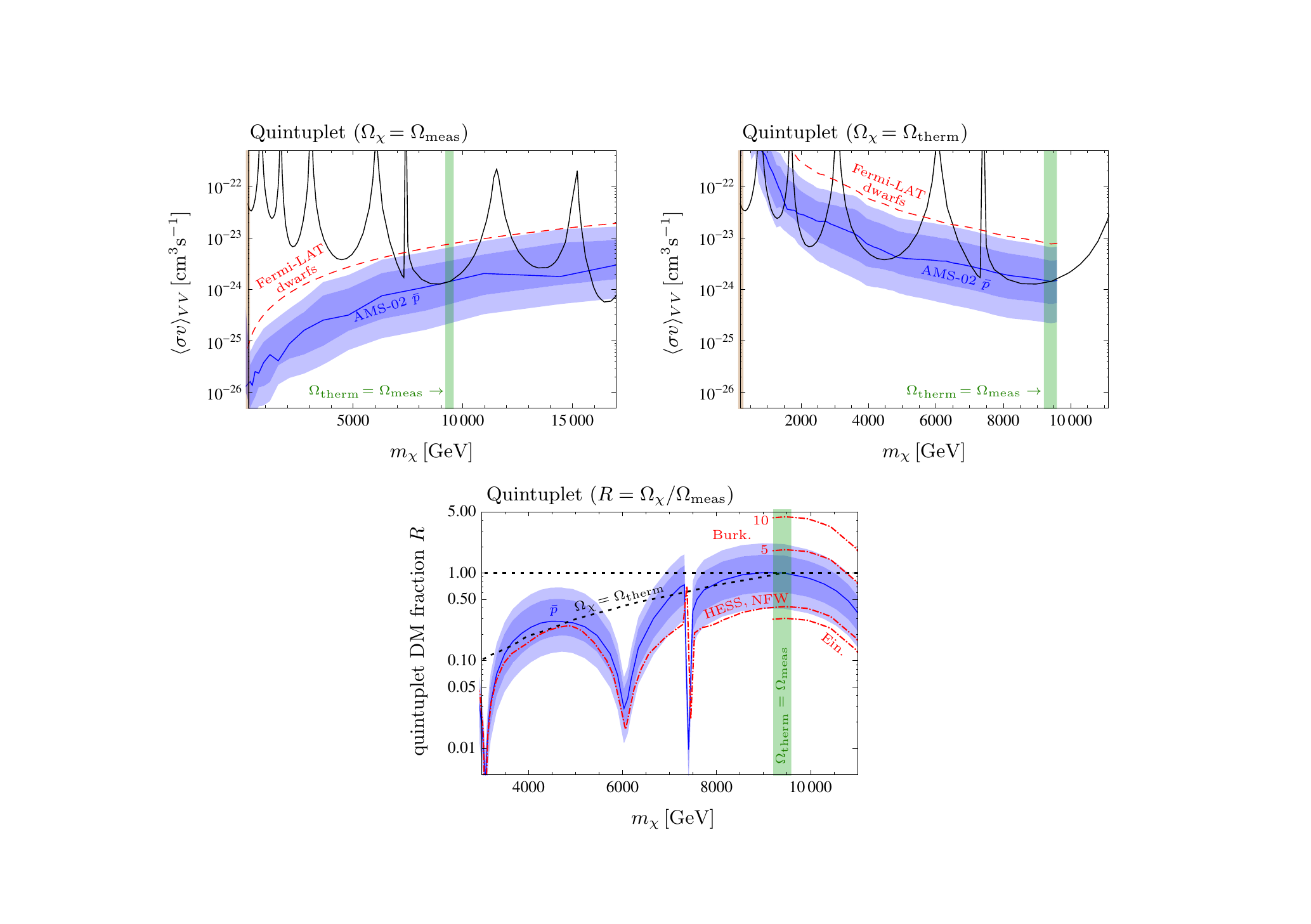}}
\end{picture}
\caption{Same as figure~\ref{fig:wino} but for quintuplet fermion dark matter.}
\label{fig:5plet}
\end{figure}

In figs.~\ref{fig:wino}, \ref{fig:higgsino} and \ref{fig:5plet} we show limits on minimal wino, higgsino and quintuplet DM, respectively, from cosmic-ray antiproton fluxes, from dwarf galaxy searches for diffuse $\gamma$-rays and from H.E.S.S.\ $\gamma$-line searches~\cite{Abramowski:2013ax}. 
The vertical green band indicates the DM mass range for which the correct thermal relic density is obtained. The position of the band is determined from comparing the predicted relic density 
with the Planck measurement $\Omega h^2 =  0.1198$~\cite{Ade:2015xua}, taking into account a relative uncertainty of the theoretical prediction of $5\%$~\cite{Cohen:2013ama}.  
We use annihilation cross sections including Sommerfeld enhancement and electroweak corrections, see
sections~\ref{sec:wino}--\ref{sec:5plet}.

In the upper left panels we show the limits on the annihilation cross section from the antiproton flux and from dwarf diffuse $\gamma$-ray searches, assuming that the minimal DM candidate constitutes all of DM so that its relic density is equal to the Planck measurement ($\Omega_\chi = \Omega_{\rm meas}$). Dark matter masses outside the green band thus correspond to scenarios with an additional (non-thermal) production mechanism or a non-standard cosmological history.\footnote{See e.g.~\cite{Fan:2013faa} for a discussion of non-thermal contributions to the wino abundance from a decay of a heavy gravitino.} The inner, dark blue band of the antiproton limit corresponds to the cosmic-ray propagation uncertainty as estimated in section~\ref{sec:limits}. As discussed there, for the mass range of interest the limits on $ZZ$ and $WW$ annihilation are virtually identical. Hence, we can apply the limits derived for $WW/ZZ$ final states to any admixture of the $WW$ and $ZZ$ channels. As the antiproton limits also depend on the local DM density, we include the corresponding uncertainty as the light blue band, which is linearly added assuming $\rho_\odot = (0.43\pm 0.15)\,$GeV/cm$^3$. 
As already pointed out in section~\ref{sec:limits}, the analysis of the antiproton flux provides a more stringent test of DM models annihilating into gauge bosons than current $\gamma$-ray searches from dwarf galaxies. 
Comparing the antiproton limit with the predicted annihilation cross section $\langle\sigma v\rangle_{VV}=\langle\sigma v\rangle_{WW}+\langle\sigma v\rangle_{ZZ}+\frac12\langle\sigma v\rangle_{Z\gamma}$ (solid black line), we can exclude minimal DM for masses below approximately 2.8, 0.3 and 7\,TeV for the case of a wino, higgsino and quintuplet, respectively, even when conservatively considering the upper edge of the uncertainty band including the local DM density error.

In the upper right panels of figs.~\ref{fig:wino}, \ref{fig:higgsino} and \ref{fig:5plet} we assume that the minimal DM relic density is equal to its thermal value ($\Omega_\chi = \Omega_{\rm therm}$). 
Dark matter masses to the left of the vertical green band result in thermal relic densities below the measured value and therefore correspond to scenarios where the minimal DM candidate does not constitute all of DM, while masses above the green band lead to an over-abundance and are, hence, excluded. In the region to the left of the vertical green band we rescale the indirect detection limits accordingly, i.e.~by the square of the ratio of minimal DM to all of DM, $R = \Omega_\chi / \Omega_{\rm meas}$. As the predicted relic density decreases with smaller masses, the resulting limits become significantly weaker for lighter DM\@. We can exclude wino DM with a thermal relic density between around 2 and 2.8\,TeV, see Fig.~\ref{fig:wino}. For a thermal quintuplet we can exclude a number of narrow windows in the DM mass, while for the thermal higgsino we cannot establish any bound from CR antiprotons, see Figs.~\ref{fig:5plet} and \ref{fig:higgsino}, respectively. The spikes in the indirect detection limits for the wino and quintuplet are due to a resonance effect in the relic density calculation including Sommerfeld corrections. 

 %=====================
%    \                                           |
%      \                                         |
%        \                                       |
%                                      \         |
%                                        \       |
%                                          \     |
%=====================

%=====================
%    \                                           |
%      \                                         |
%        \                                       |
%                                      \         |
%                                        \       |
%                                          \     |
%=====================

In the lower panels of Figs.~\ref{fig:wino}, \ref{fig:higgsino} and \ref{fig:5plet} we compare the derived limits from antiprotons to those from $\gamma$-line observations of the Galactic center. As the cross section predictions for antiproton and $\gamma$-lines are different we choose to present our results in a way that allows for a direct comparison. We provide an upper limit on the fraction of minimal DM, i.e.\ the ratio of minimal DM to all of DM, by demanding $R \le  \sqrt{\langle\sigma v\rangle_{{\rm pred.}}/\langle \sigma v\rangle_{{\rm limit}}}$.
Values of $R$ larger than one are, of course, not possible.
The spikes in the limits are again due to a resonant Sommerfeld enhancement of the annihilation cross sections. Assuming a thermal DM scenario, $\Omega_\chi = \Omega_{\rm therm}$, the DM fraction $R = \Omega_\chi / \Omega_{\rm meas}$ is a prediction of the model and shown as the black dashed line. 
The $\gamma$-line limits are displayed for four benchmark DM profiles, the Einasto and NFW profile as well as the Burkert profile with a core radius of $r_c = 5$\,kpc and $r_c = 10$\,kpc (red dot-dashed lines; from below to above). To reduce clutter we only show the result for the NFW profile over the whole mass range and restrict the other curves to the region in and above the thermal mass region. Their relative difference amounts to a constant factor. As discussed before, the $\gamma$-ray limits, which are based on searches in the central Galactic halo region, are subject to large uncertainties from the corresponding $J$-factors. In particular, $\gamma$-line limits cannot exclude any of the three cases in the thermal mass range (green bands)
if the DM profile has a sizeable core, i.e.\ for a Burkert profile with a core radius of $5$ or $10$\,kpc. 
In contrast, CR antiproton limits are more robust against uncertainties in the DM profile and, hence, provide
stronger constraints without assuming a specific profile. In particular, our results strongly disfavor a wino DM 
in the thermal mass range around 2.8\,TeV taking into account a wide range of uncertainties from systematics in CR propagation,
the DM profile and the local DM density.

Finally, we note that additional theoretical uncertainties arise from the prediction of the shape of the antiproton spectra.
As stated above, our default choice are the predictions from~\cite{Cirelli:2010xx} which include electroweak corrections in a model-independent way using electroweak splitting functions~\cite{Ciafaloni:2010ti}. For the case of wino DM we 
have compared our limits with those obtained using a model-specific prediction of the spectra 
including electroweak one-loop and Sommerfeld corrections~\cite{Hryczuk:2014hpa}. We find that using the
spectra from~\cite{Hryczuk:2014hpa} for DM masses between 1 and 3.2\,TeV our limits become stronger 
by about 40\% compared to the default ones. This provides an estimate of the size of the theoretical uncertainties arising from the prediction of the form of the antiproton spectrum; they are significantly smaller than the astrophysical uncertainties presented by the blue band displayed in the figures.

%===================================================================
\section{Conclusion}\label{sec:summary} 
%===================================================================

We have derived robust limits on dark matter (DM) annihilation in our Galaxy by analyzing the precise AMS-02 measurements of the cosmic 
ray (CR) antiproton, proton and helium fluxes.
By fitting the propagation parameters in the presence of a primary antiproton source from
DM annihilation we have explored possible correlations between a potential DM signal 
and the propagation parameters. In order to improve the coverage of the fit in the space of propagation 
parameters we have combined the parameter points that are potentially relevant for the limit setting 
from all fits performed for the various channels and propagation settings and reevaluated those points 
for the channel under consideration. In this way we have achieved a better convergence of the limits exploiting
degeneracies that we found to be present in the fit.

We have concentrated on heavy DM with masses above 200\,GeV where no hint for a potential DM
signal is present in the cosmic-ray antiproton spectrum. We have analyzed all annihilation channels into pairs of SM particles and 
derived limits on the annihilation cross section. We have explored a wide range of systematic uncertainties in the DM limits, including various propagation settings, DM profiles and antiproton cross section parametrizations.  
For the reference settings, our CR antiproton limits for DM annihilation into quarks, gluons, gauge and Higgs bosons as well as into neutrinos are stronger than those from $\gamma$-ray observations of dwarf spheroidal galaxies by roughly one order of magnitude,
and still stronger by a factor of two when taking into account the systematic uncertainties.
 For DM annihilation into charged leptons we have derived flavor-independent limits from the antiproton flux which are competitive to those from dwarfs for DM masses above about 2\,TeV. 

Finally, we have used our analysis of CR antiprotons to constrain three well-motivated minimal DM models: wino, higgsino and fermionic quintuplet DM. For those models, we have compared our limits to limits from $\gamma$-line searches in the Galactic center performed by H.E.S.S. While the limits obtained from observations of the Galactic center depend strongly on the DM profile close to the center of the Galaxy, the limits from CR antiprotons are more robust. For a cored DM density profile CR antiprotons provide significantly stronger constraints on heavy minimal DM\@. In particular, our limits strongly disfavor a thermal wino making up all of DM, regardless of the underlying DM profile.

%===================================================================
\section*{Acknowledgements}
%===================================================================

We acknowledge support by the German Research Foundation DFG through the 
research unit ``New physics at the LHC''.
Simulations were performed with computing resources granted by
RWTH Aachen University under project rwth0085.

\bibliographystyle{JHEP}
\bibliography{bibliography}

\end{document}